\documentclass[12pt,preprint]{aastex}

\begin{document}

\title{Low-Metallicity Gas Clouds in 
a Galaxy Proto-Cluster at Redshift 2.38}                                                                                                                                                                                                

\author{Paul J. Francis\altaffilmark{1}}
\affil{Research School of Astronomy and Astrophysics, the Australian
National University, Canberra 0200, Australia}
\email{pfrancis@mso.anu.edu.au}

\altaffiltext{1}{Joint Appointment with the Department of Physics, 
Faculty of Science, the Australian National University}

\and

\author{Gerard M. Williger}
\affil{Department of Physics and Astronomy, Johns Hopkins University,
3701 San Martin Drive, Baltimore, MD 21218}
\email{williger@pha.jhu.edu}

\begin{abstract}

We present high resolution spectroscopy of a QSO whose sight-line
passes through the halo of a pair of elliptical galaxies at redshift 2.38.
This pair of galaxies probably lies at the center of
a galaxy proto-cluster, and is embedded in a luminous extended Ly$\alpha$ 
nebula.

The QSO sight-line intersects two small gas clouds within this halo. 
These clouds have properties similar
to those of high velocity clouds (HVCs) seen in the halo of the Milky Way.
The gas is in a cool ($< 2 \times 10^4$K) and at least 20\% neutral phase, 
with metallicities in the
range -3.0 $<$ [Fe/H] $<$ -1.1 and neutral hydrogen column densities of
$\sim 10^{19.5}{\rm cm}^{-2}$. 

The origin of these clouds is unclear. The presence of low metallicity gas
within this possible proto-cluster implies either that the intra-cluster
medium has not been enriched with metals at this redshift, or the 
clouds are embedded within a hot, ionized, metal-rich gas phase.

\end{abstract}


\section{Introduction}

There has been considerable recent work on investigating the inter-relationship
between gas (traced by QSO absorption lines) and galaxies at high
redshifts \citep[eg.][]{pas01,fra01,wil02,ade03,zha03}. A complex picture
is emerging, with gas tracing galaxies on larger scales but some
classes of absorption-line system avoiding galaxies on small scales.

QSO absorption-line measurements such as these could potentially
shed light on the origin of the highly metal enriched intra-cluster medium 
of galaxy clusters today. Many models have been proposed 
\citep[eg.][]{ren97,chi00,mar00,agu01,del03} to produce this gas, but all have 
problems, and there is currently no consensus.

\begin{figure}
\plotone{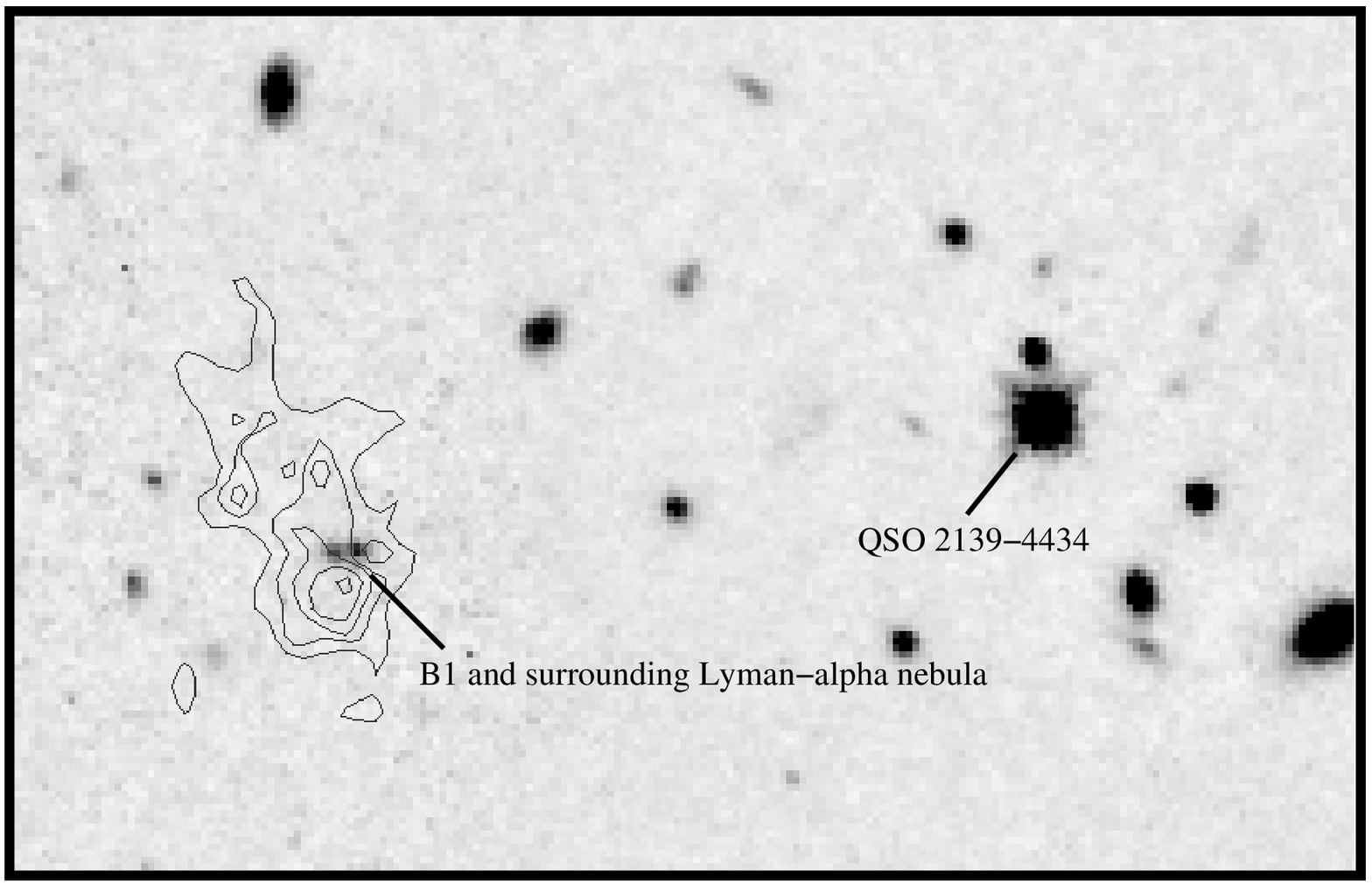}
\caption{
Close-up image of QSO 2139$-$4434, and the $z=2.38$ galaxy B1, which lies
22\arcsec\ away. Greyscale is a NICMOS F160W image (rest-frame $B$-band), 
while the contours show Ly$\alpha$ emission, measured with the Rutgers
Fabry-Perot on the CTIO Blanco Telescope.
\label{closemap}}
\end{figure}

In this paper, we present high resolution spectra of a QSO that lies behind
a probable galaxy proto-cluster at redshift 2.38 (Fig~\ref{closemap}). 
QSO LBQS 2139$-$4434 lies
at redshift 3.23. Its sight line passes only 22\arcsec\ (160 projected
proper kpc, for $H_0 = 70 
{\rm km\ s}^{-1}{\rm Mpc}^{-1}$, $\Omega_{\rm matter} = 0.3$ and 
$\Omega_{\Lambda} = 0.7$) from the extremely red object B1. \citet{fra01}
demonstrated that B1 is actually a pair of large elliptical galaxies at
redshift 2.38..

B1 is embedded within a luminous, extended Ly$\alpha$ emitting
nebula or ``blob''. Such Ly$\alpha$ nebulae are thought to 
be associated with massive dark halos, and may be the ancestors of
cooling flows today \citep{ste00,fra01}. It lies within a probable
galaxy proto-cluster \citep{fra93,fra96,fra97,fww01}, which is itself
embedded in a 80 Mpc scale ``Great Wall'' \citep{pal04,fra04}.

This sight-line thus allows us to probe the gas in an unusual part of the
early universe: the halo of an cluster elliptical galaxy.

\section{Observations and Reduction\label{obsred}}

Our existing spectra of LBQS 2139$-$4434 \citep{fra93,fww01} were of too low 
a resolution and
narrow a wavelength coverage to seriously constrain the physical properties of
the gas. We therefore re-observed it using the University College London Echelle
Spectrograph (UCLES) on the Anglo-Australian Telescope (AAT). 
Observations were carried out on the nights of 2001 August 20 --- 23 and the 
total usable integration time was 30,600 sec. \citet{dod02} independently 
obtained high resolution spectra of this QSO, 
using the UVES spectrograph. They kindly
allowed us to use their spectrum in this analysis.

The data were reduced using standard procedures, and set to the vacuum
heliocentric frame. Absorption-lines
properties were measured interactively using the XVOIGT program \citep{mar95}.
We fit our UCLES spectrum, the UVES spectrum, and a lower resolution AAT 
spectrum which gave better coverage at UV wavelengths \citep{fww01}, using
for each line whichever spectrum gave the most reliable data. Where a given
line was seen in multiple spectra, the derived parameters were always
consistent within the error bars. Parameters we derived from the UVES spectra
agree very well with those derived by \citet{dod02}.

\section{Results}

QSO LBQS 2139$-$4434 was previously known to show high equivalent width
Ly$\alpha$ absorption at the redshift of the elliptical galaxies making up
B1 (and of the cluster). We confirm this: the absorption centroid redshift
matches the \ion{C}{4} emission-line centroid of B1 to within 
$200 {\rm km\ s}^{-1}$, the uncertainty being dominated by the width of 
the emission lines.

\begin{figure}
\plotone{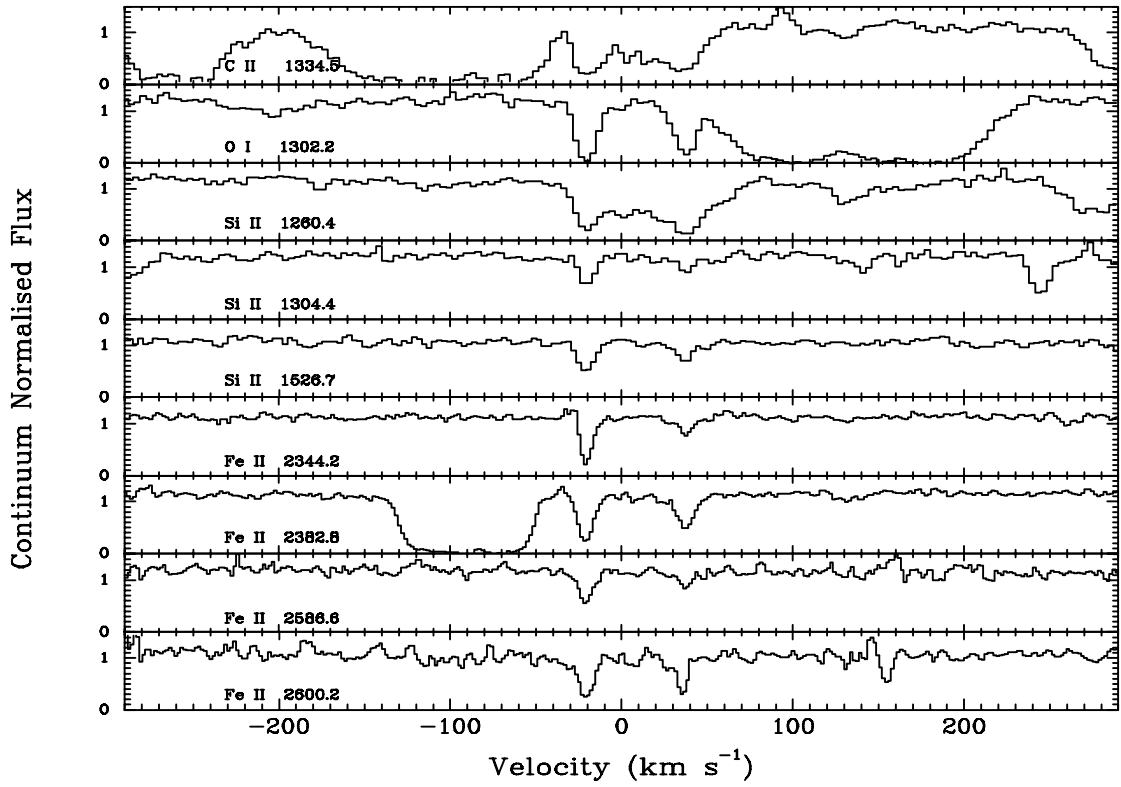}
\caption{
UVES spectra of the nine best metal-line detections in
QSO 2139-4434. Velocities are relative to a nominal 
redshift of 2.38.
\label{detplot}}
\end{figure}

For the first time, we unequivocally detect metal-line absorption from
this absorption-line system (Fig~\ref{detplot}). No high ionization
absorption is seen: the tentative detection of \ion{C}{4} by 
\citet{fww01} turns out, at higher resolution, to be caused by blends 
of unrelated lines at other redshifts.

The metal-line absorption is clearly resolved into two narrow components,
separated in velocity by $59 {\rm km\ s}^{-1}$: a lower redshift component
(component 1) at $z=2.3797$ and a higher redshift component (component 2)
at $z=2.3804$. Both components are narrow and
spectrally unresolved: we place upper limits on the Doppler $b$ parameter of 
$< 6.0 {\rm km\ s}^{-1}$ in all transitions. For the \ion{Fe}{2} 
transitions of component 1
we can place a tighter upper limit of $b < 5 {\rm km\ s}^{-1}$. There is no
evidence for velocity substructure within either of these components.

If we assume that the neutral hydrogen is located in one or both of the
metal-line clouds, this allows us to break 
the degeneracy (noted by \citet{fww01}) between velocity dispersion and 
column density in our fits to the Ly$\alpha$ emission profile, and measure
the neutral hydrogen column density. The combined neutral atomic hydrogen column
density is $10^{19.7 \pm 0.1}{\rm cm}^{-2}$. We are unable to constrain how it
is divided up between the two components.

There is also a strong degeneracy between the assumed velocity dispersion
$b$ and the inferred metal column densities. For component 1, we can place
a lower limit of $b>2 {\rm km\ s}^{-1}$ from the \ion{O}{1} line:
if $b$ were lower, then to get a good fit we would need such a high
column density that we would start to see damping wings, which are not 
observed. We cannot place an observational lower limit on $b$ in component
2, but physically the sound speed of even a cold neutral medium should be
at least $1 {\rm km\ s}^{-1}$, so we take that as our lower limit. In
Table~\ref{coldens}, we use these limits on $b$ to bracket the possible column
densities.

We list upper limits on non-detected lines by assuming the same limiting
values of $b$, and seeing interactively how strong such an absorber could be 
before the residuals go clearly into emission. 

\section{The Physical State of the Absorbing Gas}

The strength of the low ionization species, such as \ion{O}{1} and 
\ion{Si}{2}, compared with our upper limits on high ionization
species such as \ion{C}{4} and \ion{Si}{4} clearly indicate that the
absorbing gas is cool, and at least partially neutral.

There are two possibilities for the temperature of these gas clouds:
we could have a warm neutral/partially ionised phase with
$T \sim 10^4$K or a cold neutral phase with $10^2 < T < 10^3$ K
\citep[eg.][]{sut93}.

We modeled the first possibility (warm neutral/partially ionised phase) 
using Gary
Ferland's CLOUDY code \citep{fer96}. For plausible choices of the
UV ionizing background \citep{haa96,sco00}, we can reproduce the
absorption-line columns measured with a cloud of density
$0.1 < N_H < 10 {\rm cm}^{-3}$, size 2 --- 200 pc and mass 
0.001 -- 10 $M_{\sun}$. If the density is towards the lower end of this range,
the hydrogen could be up to $\sim 80$\% ionised.

The second possibility is almost impossible to model, due to the complex
molecular cooling processes, but cold neutral gas in our own galaxy does show
ionization states and column densities comparable to those we observe.

Which of these models is correct? Our stringent upper limits on the velocity
dispersions of the two components provide some evidence in favor of the cold 
model. The sound speed in the warm model is
$\sim 10 {\rm km\ s}^{-1}$, while in the cold model it would be only
1 --- 3 ${\rm km\ s}^{-1}$. One would expect the internal
turbulent velocity dispersions to be comparable to the sound speed. The
rapid movement of these clouds through B1's dark matter halo should be
ample to excite and sustain such turbulent motions. Our upper limit
of $B < 6 {\rm km\ s}^{-1}$ is thus most consistent with a cool model.

On the other hand, if the velocity dispersion really is low ($b \sim 2 {\rm km\
s}^{-1}$), we are forced to invoke extremely large \ion{O}{1} column
densities to fit the observed absorption line. This would imply
an extraordinary overabundance of oxygen compared to silicon: [O/Si] $> 1.5$
for $b \sim 2$. Larger values of $b$ bring the inferred column density of
\ion{O}{1} down to more reasonable values. A cold neutral phase with
supersonic turbulence would be an interesting possibility.

If the gas were in the cold phase, 
we might expect dust to be present. We can place upper
limits on the quantity of dust by noting, however, that the QSO has a UV
continuum slope indistinguishable from other QSOs in the Large Bright
QSO Sample \citep{fra91}. This allows us to place a rough upper limit on
the extinction in these clouds of $E(B-V)<0.1$.

We also note (Fig~\ref{detplot}) that the \ion{O}{1} line touches zero flux. 
Thus
the absorbing cloud must be large enough to completely obscure the background
QSO continuum emitting region.

In conclusion, we are unable to discriminate between the two physical
models. We consider the oxygen overabundance argument somewhat stronger
than the velocity dispersion argument, but neither is conclusive. Higher
resolution spectra will be needed to resolve this puzzle.

\section{Conclusions}

Our line of sight though the halo of B1, and probably though the central 
regions
of a galaxy proto-cluster, is thus intersecting two small gas
clouds. Their properties are very similar to those of high velocity clouds
(HVCs) in the halos of our own galaxy \citep[eg.][]{gib01,tri03}.  Their
metallicities must be low: $-3.0 < $[Fe/H] $< -1.1$,
[C/H] $< -1.65$ and $-2.9 <$ [Si/H] $<-1.65$ (Table~\ref{coldens}, and allowing
for neutral hydrogen fractions as low as 20\%). The warmer the gas, the lower
our inferred metallicity. If 
there is substantial molecular hydrogen, the metallicities would
be lower still.

Given the continued controversy over the nature even of the HVCs in our own
galactic halo, it would be foolhardy to make dogmatic statements about
the origins of these distant clouds. Many of the same models that have been 
proposed for our local HVCs (eg. tidally disrupted dwarf galaxies, in-falling 
primordial gas, gas trapped in dark matter sub-halos, debris from galactic 
fountains) could apply here.

The metallicities we measure for these clouds are lower than the typical
intra-cluster medium (ICM) metallicities of galaxy clusters today. This implies
either that this enrichment of this cluster had not yet taken place, or
that the cold gas clouds we see are in some way isolated from the enriched
gas. It is possible, for example, that these clouds are embedded in a very hot,
ionized, X-ray emitting, metal-rich ICM. We would not detect such a hot phase in
absorption. We do not detect B1 in X-rays (Williger et al. 2004, in
preparation), but our current limits do not place tight constraints on the
existence of such an ICM.

These observations may also shed light on the enigmatic Ly$\alpha$ blobs. While
our sight-line does not pass through the blob, it does probe the same dark
matter halo. It is quite possible that the Ly$\alpha$ emission region is made up
of small clouds similar to those we are probing in absorption. The emission
could be driven by photoionization and photo-evaporation of these clouds by
some concealed AGN, or could be produced by fast shocks created at cloud-cloud
collisions, as discussed by \citet{fra01}.

\acknowledgments

We would like to thank D'Odorico, Petitjean and Cristiani for allowing us to
use their UVES spectrum of QSO 2139$-$4434, and Ralph Sutherland for
many illuminating discussions. We would also like to thank all
the people who helped us get back to work at Mt Stromlo Observatory so soon
after the January 2003 bushfires.

\clearpage


\clearpage

\clearpage

\begin{deluxetable}{lcccc}
\tablewidth{0pt}
\tablecolumns{5}
\tablecaption{LBQS 2139$-$4434 Absorption-Line Column Densities \label{coldens}}
\tablehead{
\colhead{Ion} & 
\multicolumn{4}{c}{Log(Column Density) (cm$^{-2}$)} \\
\colhead{ ~ }  & \multicolumn{2}{c}{z=2.3797 Component} & 
\multicolumn{2}{c}{z=2.3804 Component} \\

\colhead{ ~ } & \colhead{$b=2{\rm km\ s}^{-1}$} & 
\colhead{$b=5{\rm km\ s}^{-1}$} &
\colhead{$b=1{\rm km\ s}^{-1}$} &
\colhead{$b=6{\rm km\ s}^{-1}$} \\
}
\startdata

\ion{H}{1}  &    \multicolumn{4}{c}{19.6 --- 19.8} \\
\ion{C}{1}  & $<$12.2 & $<$12.2 & $<$12.8 & $<$12.8 \\
\ion{C}{2}  & $<$14.5 & $<$15.5 & $<$14.4 & $<$13.9 \\
\ion{C}{4}  & $<$12.4 & $<$12.5 & $<$12.5 & $<$12.5 \\
\ion{O}{1}  
    & 16.4 --- 17.0 & 14.2 --- 14.7 & 16.9 --- 17.3 & 14.0 --- 14.3 \\
\ion{Si}{1} & $<$11.7 & $<$11.9 & $<$11.6 & $<$11.9 \\
\ion{Si}{2} 
    & 13.0 --- 13.3 & 13.0 --- 13.1 & 12.4 --- 13.1 & 12.5 --- 12.9 \\
\ion{Si}{4} & $<$11.6 & $<$11.8 & $<$12.2 & $<$12.4 \\
\ion{S}{1}  & $<$12.0 & $<$11.9 & $<$12.4 & $<$12.6 \\
\ion{Fe}{2} 
    & 13.8 --- 13.9 & 13.0 --- 13.1 & 13.0 --- 13.2 & 12.6 --- 12.8 \\

\enddata
\end{deluxetable}

\end{document}